\documentclass[10pt]{iopart}
\usepackage{iopams}

\usepackage{geometry}
\usepackage{graphicx}
\usepackage{bm}
\usepackage[urlcolor=blue,colorlinks=true,citecolor=blue,linkcolor=blue,pdfstartview={FitH},bookmarks=false]{hyperref}
\usepackage{amssymb}
\usepackage{xcolor}
\usepackage{epstopdf}

\begin{document}

\title[Probe-type of superconductivity by impurity in materials with short coherence length]{Probe-type of superconductivity by impurity in materials with short coherence length: the {\it s-wave} and $\eta$-{\it wave} phases study}

\author{Andrzej Ptok$^{1}$, Konrad Jerzy Kapcia$^{2,3}$}
\address{$^1$ Institute of Physics, University of Silesia, 40-007 Katowice, Poland}
\address{$^2$ Electron States of Solids Division, Faculty of Physics, Adam Mickiewicz University in Pozna\'{n}, Umultowska 85, 61-614 Pozna\'{n}, Poland}
\address{$^3$ Condensed Matter Theory Sector, International School for Advanced Studies (SISSA), via Bonomea 265, 34136, Trieste, Italy}
\eads{\mailto{aptok@mmj.pl}, \mailto{konrad.kapcia@amu.edu.pl}}
\vspace{10pt}

\begin{indented}
\item[]August 2014
\end{indented}

\begin{abstract}
The effects of a single non-magnetic impurity on superconducting states in the Penson-Kolb-Hubbard model have been  analyzed.
The investigations have been performed within two steps: (i) the homogeneous system is analysed using the Bogoliubov transformation, whereas (ii) the inhomogeneous system is investigated self-consistent Bogoliubova-de Gennes equations (by exact diagonalization and the kernel polynomial method).
We analysed both signs of pair hopping, which correspond to {\it s-wave} and $\eta${\it-wave} superconductivity.
Our results show that an enhancement of the local superconducting gap at the impurity-site occurs for both cases.
We obtained that Cooper pairs are scattered (at the impurity site) into the states which are from the neighborhoods of the states, which are commensurate ones with the crystal lattice.
Additionally, in the $\eta$-phase there are peaks in the local-energy gap (in momentum space), which are connected with long-range oscillations in the spatial distribution of the energy gap, superconducting order parameter as well as effective pairing potential. Our results can be contrasted with the experiment and predicts how to experimentally differentiate these two different symmetries of superconducting order parameter by scanning tunneling microscopy technique.
\end{abstract}


\vspace{2pc}

\vspace*{2pc}

\noindent (Some figures may appear in colour only in the online journal)

\submitto{\SUST}

\maketitle
%
%


\section{Introduction}

The Penson-Kolb-Hubbard (PKH) model is one of the conceptually simplest phenomenological models for studying correlations and for description of superconductivity (SC) in narrow-band systems with short-range, almost unretarded pairing. Generally, impurities destroy or worsen desirable SC properties, but sometimes can induce new interesting phenomena.

The cuprates and iron-pnictides are examples of superconductors with short coherence length. Moreover, such systems are highly inhomogeneous and the role of impurities on SC could be crucial. In both groups of materials the superconducting state itself is generated by chemical doping which inevitably disorders the samples, and second, local probes of the quasi-particle states near the impurity sites can provide important information on the underlying system~\cite{balatsky.vekhter.06,alloul.bobroff.09}. In discussed materials the physical properties changing by doping, the result of which are phase transition. The phase diagrams for these materials include antiferromagnetically ordered (AF), metallic (non-ordered) and superconducting (SC) phases. With increasing concentration of current carriers  the AF phase is vanishing first, then the SC phase occurs in define range of doping~\cite{johnston.10,micnas.ranninger.90}. The other interesting feature of these compounds are effects introduced by (nonmagnetic) impurities such as  e.g. inhomogeneity superconducting state~\cite{maska.sledz.07,krzyszczak.domanski.10,mashima.fukuo.06}, pining of vortexes~\cite{mashima.fukuo.06,fukuo.mashima.06}, modification persistent current in superconducting ring~\cite{czajka.maska.05}, and induced spin density waves~\cite{kim.chen.09}. Notice that the effects of nonmagnetic impurities in  above materials are qualitatively different than in conventional superconductors, any perturbation that does not lift the Kramers degeneracy of these states does not affect the mean-field superconducting transition temperature.

The pair-hopping term ($J$) was proposed in Ref.~\cite{penson.kolb.86,kolb.penson.86} and can be derived from a general microscopic tight-binding Hamiltonian, where the Coulomb repulsion may lead to the pair hopping interaction $J =  \langle ii | e^{2} / r | jj \rangle$~\cite{czart.robaszkiewicz.01,robaszkiewicz.czart.03,czart.robaszkiewicz.07}. In such a case $J$ is positive (\emph{repulsive} model $J > 0$, favoring $\eta${\it-wave} SC), but in this case the magnitude of $J$ is very small. However, the  effective attractive form ($J<0$, favoring {\it s-wave} SC) is also possible (as well as an enhancement of the magnitude of $J>0$) and it can originate from the coupling of electrons with intersite (intermolecular) vibrations via modulation of the hopping integral or from the on-site hybridization term in the general periodic Anderson model (cf. e.g.~\cite{micnas.ranninger.90,robaszkiewicz.bulka.99,kapcia.robaszkiewicz.13,kapcia.robaszkiewicz.12,mierzejewski.maska.04} and references therein). It can also be included in  the effective models for Fermi gas in an optical lattice in the strong interaction limit~\cite{rosch.rasch.08}. The role of $J$ interaction in a multiorbital model is of a particular interest because of its presence in the iron pnictides~\cite{ptok.kapcia.14}. Notice that it has been found that SC originating from the \mbox{$J>0$} interaction is unique in that it is robust against the orbital (diamagnetic) pair breaking mechanism~\cite{mierzejewski.maska.04}.

In this paper we investigate a role of single nonmagnetic impurity on SC states ({\it s-wave} as well as $\eta${\it-wave}). We obtain the spatial distribution of local energy gap (and its Fourier transform), superconducting order parameter as well as effective pairing potential. We also propose an experimental method that can determine what type of superconductivity ({\it s-wave} or $\eta${\it-wave}) occurs in the material.

\section{Model and technique}

We explore two dimensional square lattice of a size $N_x \times N_y$ with the periodic boundary conditions, where we assume the possibility of the hopping of {\it (i)} electrons (with amplitude $t$) and {\it (ii)} pairs of electrons (with amplitude $J$) between nearest neighbors (NN). Electrons with opposite spin at the same site interacts with energy $U$ (inter-site Coulomb interaction). In the model we also introduce the diagonal disorder induced by non-magnetic impurities.
The Hamiltonian of the system in real space can be describe as $H = H_{0} + H_{int}$, where:
\begin{eqnarray}
H_{0} &=& \sum_{ \langle i,j \rangle \sigma } \left\lbrace - t + \left( V_{i} - \mu ) \right) \delta_{ij} \right\rbrace c_{i\sigma}^{\dagger} c_{j\sigma} , \\
\label{eq.ham_int} H_{int} &=& U \sum_{i} c_{i\uparrow}^{\dagger} c_{i\downarrow}^{\dagger} c_{i\downarrow} c_{i\uparrow} + J \sum_{ \langle i,j \rangle } c_{i\uparrow}^{\dagger} c_{i\downarrow}^{\dagger} c_{j\downarrow} c_{j\uparrow} .
\end{eqnarray}
$\sum_{\langle i,j \rangle}$ restricts the summation to NN, $c_{i\sigma}$ ($c_{i\sigma}^{\dagger}$) are annihilation (creation) operators of electron at $i$-th site  with spin $\sigma = \{ \uparrow , \downarrow \}$, $\mu$ is chemical potential and $V_{i}$ is non-magnetic impurity potential at $i$-th site. The electron hopping amplitude ($t$) will be taken as a scale of energy in the system.

\begin{figure}[!ht]
\begin{center}
\includegraphics[bb=0 0 0 80]{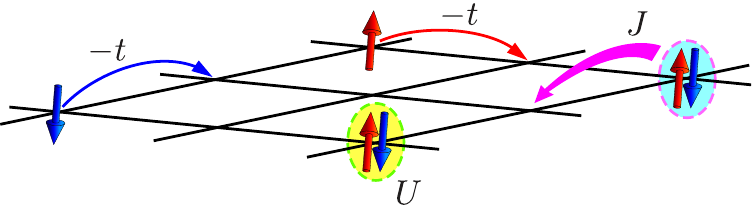}
\end{center}
\caption{
Schematic illustration of the model analyzed.
The impurity can be identified as a change of a local potential at the impurity site (not shown explicitly).
}
\label{fig.schemat}
\end{figure}

In the mean-field Hartree-Fock approximation \cite{micnas.ranninger.90} interaction Hamiltonian~(\ref{eq.ham_int}) takes a form:
\begin{eqnarray}
H^{HF}_{int} &=& U \sum_{i} \left( \Delta_{i}^{\ast} c_{i\downarrow} c_{i\uparrow} + H.c. 
\right) - U \sum_{i} | \Delta_{i} |^{2} \\
\nonumber &+& J \sum_{ \langle i,j \rangle } \left( \Delta^{\ast}_{j} c_{i\downarrow} c_{i\uparrow} + H.c. 
\right) - J \sum_{ \langle i,j \rangle } \Delta^{\ast}_{i}  \Delta_{j}
\end{eqnarray}
where we introduce dimensionless superconducting order parameters (SOP) $\Delta_{i} = \langle c_{i\downarrow}c_{i\uparrow} \rangle$.
Thus, in the non-homogeneous at every site the effective (local) energy gap exists, which value is determined by
\begin{eqnarray}\label{eq.energy_gap}
\mathcal{D}_{i} &=& U \Delta_{i} + J \sum_{ \langle i,j \rangle } \Delta_{j} .
\end{eqnarray}
This quantity is connected with the effective pairing potential defined as $U^{eff}_{i} = \mathcal{D}_{i} / | \Delta_{i} |$, which is also not a constant value in a heterogeneous system.
To analyze the gap distribution in the momentum space we introduce also its Fourier transform (FT) defined as
\begin{eqnarray}\label{eq.fft}
\mathcal{D}_{\bm k} &=& \frac{1}{N_{x} N_{y}} \sum_{i} \mathcal{D}_{i} \exp ( - i {\bm k} \cdot {\bm r}_{i} ) .
\end{eqnarray}
${\bm k}$ denote a momentum of Cooper pair.

The further analyses are performed in two steps: {\it (i)} the homogeneous system is analyzed using the Bogoliubov transformation, whereas {\it (ii)} the inhomogeneous system is investigated by self-consistent Bogoliubova-de Gennes equations in real space approach~\cite{bdge}, (by exact diagonalization and the kernel polynomial method~\cite{weisse.wellein.06}).
The both approaches will be described briefly below.

\subsection{Homogeneous system}

\begin{figure}[!ht]
\begin{center}
\includegraphics[scale=1.5,bb=0 0 0 140]{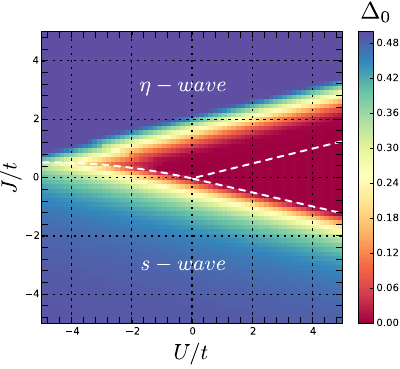}
\end{center}
\caption{
The ground state phase diagram of the model in homogeneous system ($V_{i}=0.0t$ and $\mu = 0.0t$). White dashed lines show (indicatively) boundaries of the phases.} 
\label{fig.df}
\end{figure}

In a general case the SC state is characterized by the formation of the Cooper pairs with total momentum $\bm Q$.
In a case ${\bm Q} \neq \bm 0$ the SC state will be called Fulde-Ferrell-Larkin-Ovchinnikov (FFLO) state~\cite{FF,LO,matsuda.shimahara.07}.
Notice that $\bm Q = \bm 0 \equiv(0,0)$ case (favouring by $J<0$) corresponds to {\it s-wave} SC (BCS-like SC), whereas  ${\bm Q}={\bf \Pi}\equiv(\pi,\pi)$ (favouring by $J>0$) corresponds to $\eta${\it-wave} SC~\cite{robaszkiewicz.bulka.99}, in which the SOP alternates
from one site to the neighbouring one.
In particular, for square lattice one can distinguish two sublattices in which the phase of SOP will differ by $\pi$, whereas e.g. for triangular lattice one can distinguish three sublattices in which the phase of SOP will differ by $2\pi/3$~\cite{ptok.mierzejewski.08}.
In the homogeneous system only the component $\mathcal{D}_{\bm k}$ (Eq.~(\ref{eq.fft})) with ${\bm k } = {\bm 0}$ or ${\bm k} = {\bm \Pi}$ is nonzero ({\it s}- or $\eta${\it-wave}, respectively).
Thus, in homogeneous system ($V_{i} = 0$) the SOP can be derived as $\Delta_{i} = \Delta_{0} \exp ( i {\bm Q} \cdot {\bm r}_{i} ) $.
Moreover, it has been shown that flux quantization and the Meissner effect appear in {\it s-wave} as well as in $\eta${\it-wave} SC state~\cite{yang.62}.
The mean-field Hamiltonian $H^{HF} = H_{0} + H^{HF}_{int}$  in the momentum space takes the form~\cite{ptok.maska.09}:
\begin{eqnarray}
H^{HF} &=& \sum_{{\bm k} \sigma } \left\lbrace - t \gamma_{\bm k} - \mu \right\rbrace c_{{\bm k}\sigma}^{\dagger} c_{{\bm k}\sigma} \\
\nonumber &+& U_{\bm Q}^{eff} \sum_{\bm k} \left( \Delta_{0}^{\ast} c_{-{\bm k}+{\bm Q} \downarrow} c_{{\bm k}\uparrow} + H.c. \right) - U_{\bm Q}^{eff} N | \Delta_{0} |^{2}
\end{eqnarray}
where $\gamma_{\bm k} = 2 \left ( \cos ( k_{x} ) + \cos ( k_{y} ) \right)$,  $U_{\bm k}^{eff} = U + J \gamma_{\bm k}$ is a~effective pairing interaction in the momentum space, and $N=N_{x} \times N_{y}$. Moreover, in this case $U_{\bm k}^{eff} = U_{i}^{eff}$.
Using the Bogoliubov transformation~\cite{ptok.crivelli.13,ptok.14} one can obtain a spectrum of the Hamiltonian as
\begin{eqnarray}
\mathcal{E}_{{\bm k},\pm} &=& \left( E_{{\bm k}\uparrow} - E_{-{\bm k}+{\bm Q}\downarrow} \right) / 2 \\
\nonumber &\pm & \sqrt{ \left( E_{{\bm k}\uparrow} + E_{-{\bm k}+{\bm Q}\downarrow} \right)^{2} / 4 + | U_{\bm Q}^{eff} \Delta_{0} |^{2} }
\end{eqnarray}
Straightforward calculations lead to the following form of the grand canonical potential $\Omega = -k_BT \ln {\rm Tr}[ \exp(-\beta H^{HF})]$:
\begin{eqnarray}
\label{eq.free_ene} \Omega &=& - k_{B} T \sum_{{\bm k},\tau=\pm} \ln \left( 1 + \exp ( - \beta \mathcal{E}_{{\bm k}\tau} ) \right) \\
\nonumber &-& U_{\bm Q}^{eff} N | \Delta_{0} |^{2} + \sum_{\bm k} E_{{\bm k},-}
\end{eqnarray}
where $\beta = 1 / k_{B} T$. Notice that Eq.~(\ref{eq.free_ene}) is also valid for the inhomogeneous system, but in that case the Hamiltonian spectrum can be determined only numerically.
The ground state for fixed model parameters is founded by a minimization of $\Omega$ with respect to $\Delta_0$. It enables, among other,  a founding the ground state phase diagram (cf. Fig.~\ref{fig.df}), from which we choose values of model parameters to further numerical calculation in inhomogeneous system.

\subsection{Inhomogeneous system}

The superconducting states in an inhomogeneous system are described by the following mean-field single particle Hamiltonian written in Nambu space:
\begin{eqnarray}
\label{eq.ham_nambu}\label{eq.ham_mfa} H = \sum_{ \langle i,j \rangle } \left( \begin{array}{cc} c_{i\uparrow}^{\dagger} & c_{i\downarrow} \end{array} \right)
\left( \begin{array}{cc}
H_{ij\uparrow} & \Delta_{ij} \\
\Delta_{ij}^{\ast} & -H_{ij\downarrow}^{\ast}
\end{array} \right) \left( \begin{array}{c} c_{j\uparrow} \\ c_{j\downarrow}^{\dagger} \end{array} \right) .
\end{eqnarray}
The off-diagonal elements $\Delta_{ij} = \delta_{ij} ( U \Delta_{i} + J \sum_{ \langle i,k \rangle } \Delta_{k} )$ are the on-site SOP at the site {\it i}, while the diagonal elements $H_{ij\sigma} = - t_{ij} + \left( V_{i} - \mu ) \right) \delta_{ij}$ are the single particle Hamiltonian in the presence of a~given configuration of impurities.

The inhomogeneous system described by Eq.~(\ref{eq.ham_nambu}) can be solve using the Bogoliubov-Valatin transformation:
\begin{eqnarray}
c_{i\sigma} = \sum_{n} \left( u_{i n\sigma} \gamma_{n\sigma} - \sigma v_{i n \sigma}^{\ast} \gamma_{n \bar{\sigma}}^{\dagger} \right) ,
\end{eqnarray}
where $\gamma_{n\sigma}$ and $\gamma_{n\sigma}^{\dagger}$ are the quasi-particle operators, $u_{kn\sigma}$ and $v_{kn\sigma}$ are the Bogoliubov--de Gennes (BdG) eigenvectors. One can obtain the self-consistent BdG equations in real space:
\begin{eqnarray}\label{eq.bdg}
\mathcal{E}_{n \sigma} \left( \begin{array}{c}
u_{in\sigma} \\
v_{in\bar{\sigma}}
\end{array} \right) = \sum_{j} \left( \begin{array}{cc}
H_{ij\sigma} & \Delta_{ij} \\
\Delta_{ij}^{\ast} & -H_{ij\bar{\sigma}}^{\ast}
\end{array} \right) \left( \begin{array}{c}
u_{jn\sigma} \\
v_{jn\bar{\sigma}}
\end{array} \right) ,
\end{eqnarray}
where $H_{ij\sigma}$ and $\Delta_{ij}$ have been defined previously. The on-site SOP 
are given by:
\begin{eqnarray}
\Delta_{i} &=& \langle c_{i\downarrow}c_{i\uparrow} \rangle \\
\nonumber &=& \sum_{n} \left( u_{in\uparrow} v_{in\downarrow}^{\ast} f(\mathcal{E}_{n\uparrow}) - u_{in\downarrow} v_{in\uparrow}^{\ast} f(-\mathcal{E}_{n\downarrow}) \right) ,
\end{eqnarray}
where $f(E) = 1 / \left( \exp (\beta E) +1 \right)$ is the Fermi-Dirac distribution function.

This method enables to determine the SOP in inhomogeneous system in self-consistent way~\cite{maska.sledz.07,krzyszczak.domanski.10,czajka.maska.05,ptok.10,wang.hu.07,sledz.mierzejewski.06,mierzejewski.ptok.09,qiang.10,ptok.maska.11,ptok.12}.
Because it is necessary to solve eigenproblem~(\ref{eq.bdg}) of the Hamiltonian, the method can be used for systems of a size of the order $40 \times 40$.
Thus we also implement kernel polynomial method~\cite{weisse.wellein.06}, which is the extension of Bogoliubov-de Gennes equations in Chebyshev polynomial basis~\cite{alvarez.schulthess.06,furukawa.motome.04,covaci.peeters.10,gao.huang.12,nagai.nakai.12,nagai.ota.12,nagai.shinohara.13,he.song.13}.
Such a method is iterative and enables  to determined SOP in systems, which size is limited by computing time. Below we briefly discuss the main ideas of this approach.

\begin{figure}[!tbh]
\begin{center}
\includegraphics[scale=0.39,bb=0 0 360 1350]{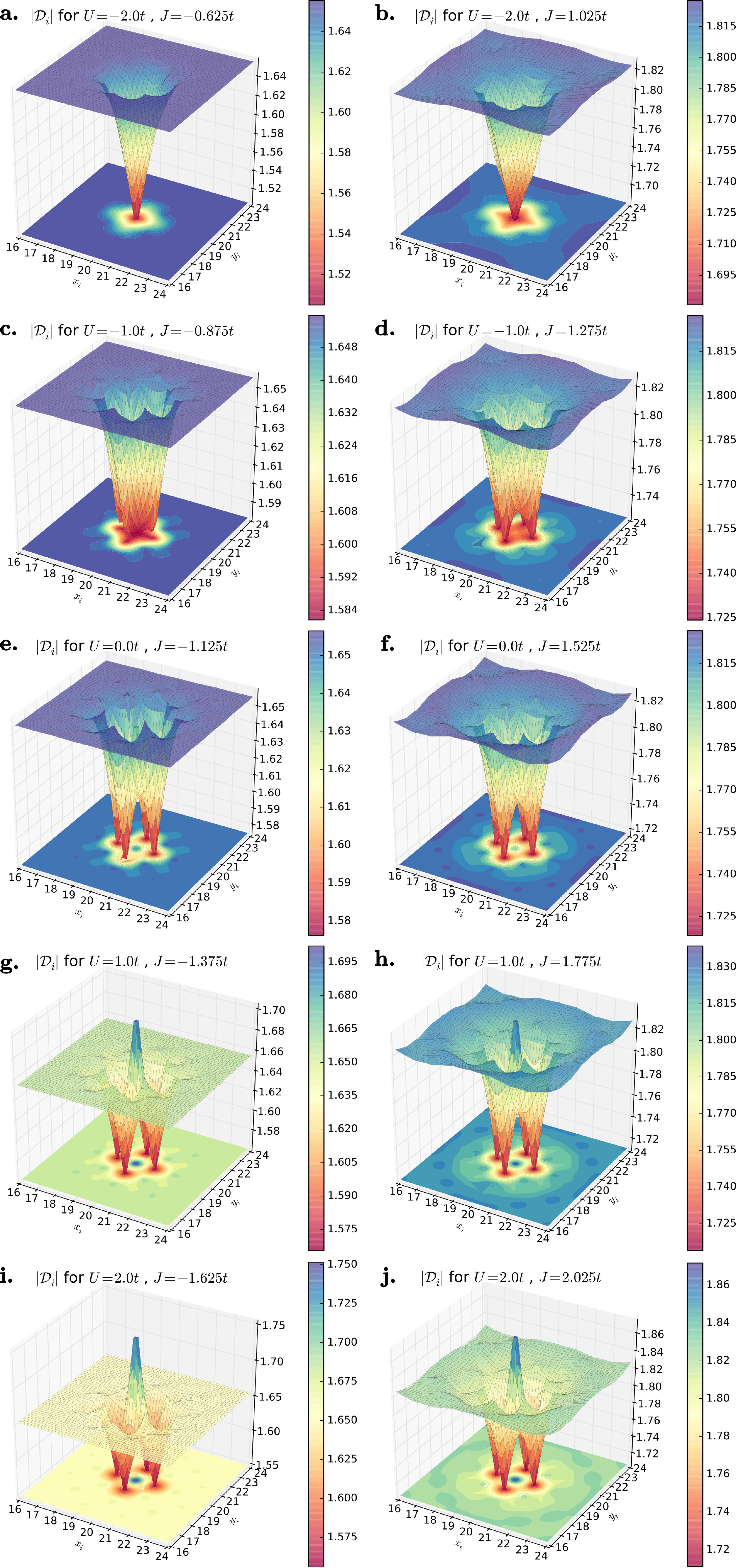}
\caption{
Energy gap $|\mathcal{D}_{i}|$ for chosen $U$ and $J$ in the {\it s-wave} and $\eta${\it-wave} phases (left and right panels, respectively) for $\mu = 0.0 t$ and $V_{imp} = 2.0 t$ (the system of a size $40\times40$).
}\label{fig.gapall}
\end{center}
\end{figure}

The Chebyshev polynomials can be written as $\phi_{k} (x) = \cos [ k \arccos x ]$ and satisfy recurrence relation:
\begin{eqnarray}
\phi_{0} (x) = 1, \quad \phi_{1} (x) = x , \\
\nonumber \phi_{k+1} (x) = 2 x \phi_{k} (x) - \phi_{k-1} (x) , \quad (k>1).
\end{eqnarray}
The polynomials $\phi_{k}$ ($k = 0,1,\ldots$) are orthonormal basis with interval $[-1,1]$:
\begin{eqnarray}
\delta ( x - x' ) = \sum_{n=0}^{\infty} \frac{W(x)}{w_{k}} \phi_{k} (x) \phi_{k} (x'), \quad x\in [-1,1] \\
W(x) = \frac{1}{\sqrt{1-x^{2}}} , \quad w_{k} = \frac{\pi(1-\delta_{k0})}{2} .
\end{eqnarray}
To obtain eigenvectors of Hamiltonian~(\ref{eq.ham_mfa}) in the basis of function $\phi_{k}$ one need to scale its matrix representation:
\begin{eqnarray}
H' = (H - b I)/ a , \quad \varepsilon_{\gamma} = (\mathcal{E}_{\gamma} - b)/a \\
\nonumber a = ( \mathcal{E}_{max} - \mathcal{E}_{min} )/2 , \quad b = (\mathcal{E}_{max} + \mathcal{E}_{min} )/2,
\end{eqnarray}
where $\mathcal{E}_{min} \leq \mathcal{E}_{\gamma} \leq \mathcal{E}_{max}$ is the range of eigenenergies of hamiltonian $H$, $H'$ is a scaled Hamiltonian with (also scaled) dimensionless eigenvalues $\varepsilon_{\gamma} \in [-1,1]$. Using the constant $2 N_{x} N_{y}$-dimensional real vectors  $[ {\bm e} (i) ]_{\gamma} = \delta_{i,\gamma}$ and $[ {\bm h} (i) ]_{\gamma} = \delta_{i+N_{x}N_{y},\gamma}$, we obtain
\begin{eqnarray}
\label{eq.chyb2}
\Delta_{i} = \langle c_{i\downarrow}c_{i\uparrow} \rangle &=& \sum_{n=0}^{\infty} {\bm e}^{T} (i) {\bm h}_{n} (i) \frac{T_{n}}{w_{n}}
\end{eqnarray}
where
\begin{eqnarray}
T_{n} = \int_{-1}^{1} dx f(ax+b) W(x) \phi_{n} (x) , \\
\nonumber {\bm e}_{n} (i) = \phi_{n} ( H' ) {\bm e} (i),  \quad {\bm h}_{n} (i) = \phi_{n} ( H' ) {\bm h} (i) .
\end{eqnarray}
A sequence of the vector ${\bm q}_{n} \equiv \phi_{n} ( H' ) {\bm q}$ (${\bm q} = {\bm e} ( i) , {\bm h} ( i )$) is recursively generated by:
\begin{eqnarray}
{\bm q}_{n+1} = 2 H' {\bm q}_{n} - {\bm q}_{n-1} \quad ( n \geq 2 ) , \\
\nonumber {\bm q}_{1} = \phi_{1} ( H' ) {\bm q} , \quad {\bm q}_{0} ( H' ) = {\bm q} .
\end{eqnarray}
$T_{n}$ does not depend on the index {\it i}. Therefore, the calculation of $T_{n}$ can be done before any other calculations. The use of the recurrence formula leads to a self-consistent calculation of the BdG equations, without any diagonalization of $H$.

The sum in (\ref{eq.chyb2}) is truncated to a certain cutoff $M$ and such an abrupt truncation generally results in unwanted Gibbs oscillations. For this reason all coefficients $T_{n}$ must be multiplied by damping factors $g_{n}$. A possible choice of the damping factors can be found in Ref.~\cite{weisse.wellein.06}.

\section{Results and discussion}

We analyze the two dimensional square lattice with the periodic boundary conditions at $kT = 10^{-5} t$ and $\mu = 0.0t$ (it corresponds to electron concentration $n = 1$ independently of the other model parameters).

\subsection{Homogeneous system}
Using the mean-field approach described before, we obtain the $J/t$ vs. $U/t$ phase diagram for homogeneous system ($V_{i}=0$) of a size $600\times600$ shown in Fig.~\ref{fig.df}.
The diagram is nonsymmetric  with respect to $J=0$ and consists of three regions, separated by dashed lines, in which different phases occur.
The $\eta${\it-wave} SC  can occur only for $J>0$, whereas {\it s-wave} phase can be stable for $J<0$ as well as for $J>0$ (in restricted range).
A necessary condition for SC phases occurrence is $U^{eff}_{i} <0$ (or $U^{eff}_{\bm k} < 0$), thus the regions of SC phases must be restricted at least by lines $U \pm 4 J = 0$.
However, the SC ({\it s}- or $\eta${\it-wave}) phase is stable only if $|U^{eff}_i|$ is higher than some critical value and the boundary of stability of the particular phases determined by minimization of $\Omega$, denoted by dashed line on the diagram, are moved towards these lines such as shown in Fig.~\ref{fig.df}. On the diagram the value of SOP $|\Delta_0|$ is also shown.
It should be stressed that computations in the real space and the momentum spaces give the same results.

\subsection{Inhomogeneous system}
Next we analyze the system of the size $40\times40$ with a single impurity $V_i = V_{imp}=2.0t$ located at the center site of the system with coordinates $(N_x/2,N_y/2)$ ($V_i=0$ at other sites). In Figs.~\ref{fig.gapall} there are shown the dependencies (as a function of lattice site) of local effective energy gap $\mathcal{D}_{i}$ (Eq.~\ref{eq.energy_gap}). The panels in these figures, respectively for {\it s-wave} and {$\eta${\it-wave} SC,  are obtained for the same (approximately) distance from the boundaries with non-ordered phase shown (for a given phase) in Fig.~\ref{fig.df} (denoted as dashed lines). It corresponds to constant value of $U_{eff}$ at a particular phase.

The $|\mathcal{D}_{imp}|$ ($|\mathcal{D}_{i}|$ at the impurity site)  initially for attractive $U$ is smaller than $|\mathcal{D}_{\infty}|$ ($|\mathcal{D}_{i}|$ far from the impurity). It increases with increasing $U$ and finally  $|\mathcal{D}_{imp}|> |\mathcal{D}_{\infty}|$ for repulsive $U$.
The valley at the impurity site vanishes faster in {\it s-wave} phase, whereas in the $\eta${\it-wave} phase it extends over a larger space (over larger number of sites).
Notice also that spatial variability of $|\mathcal{D}_{i}|$ is bigger in $\eta${\it-wave} phase than in {\it s-wave} phase (cf. also Fig.~\ref{fig.bigg} for the system size of $150\times 150$). These effects are better visible for the larger sizes of the system and is discussed in details below.

\begin{figure}[!t]
\begin{center}
\includegraphics[scale=0.35,bb=0 0 360 360]{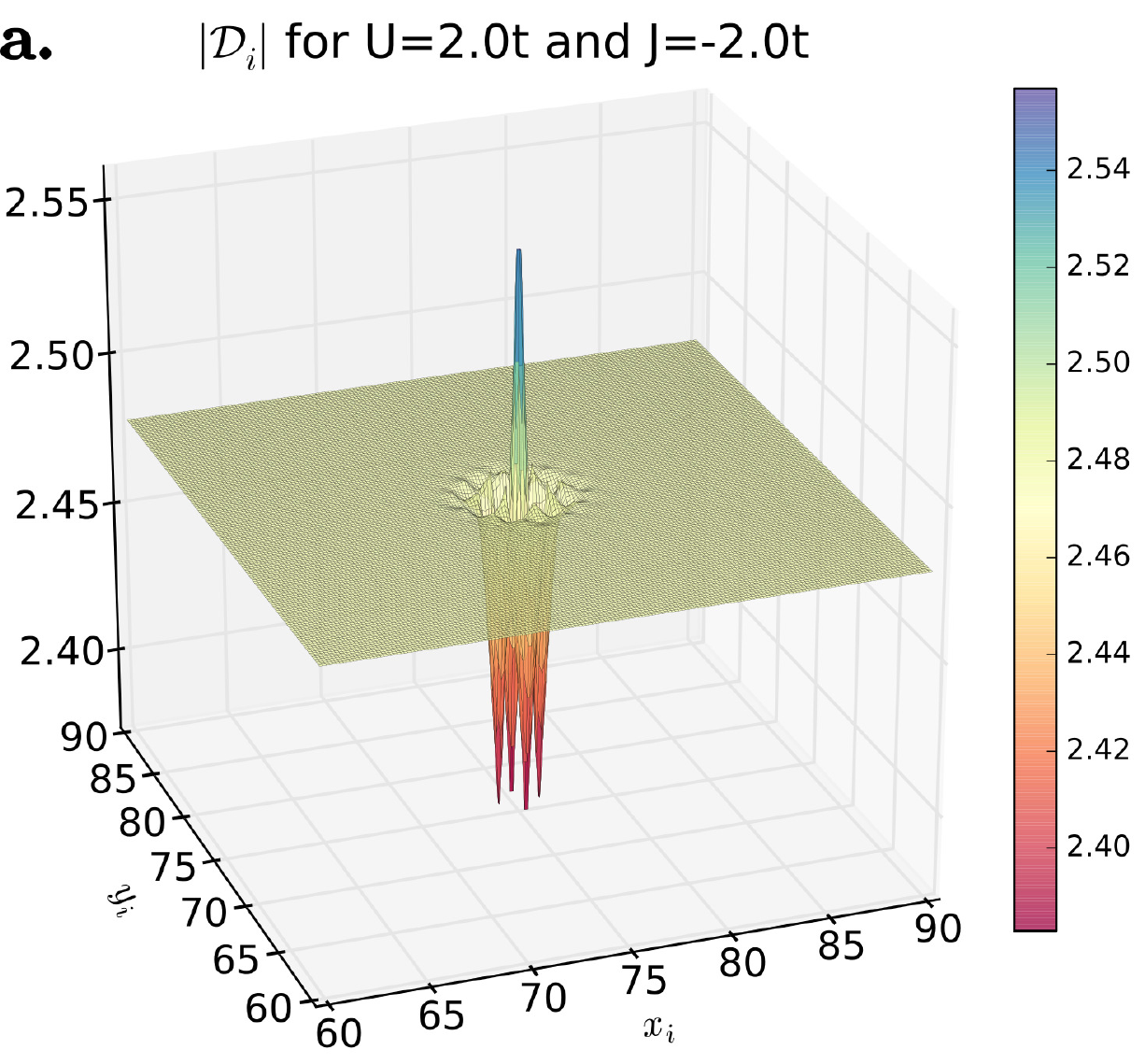}
\includegraphics[scale=0.35,bb=0 0 360 360]{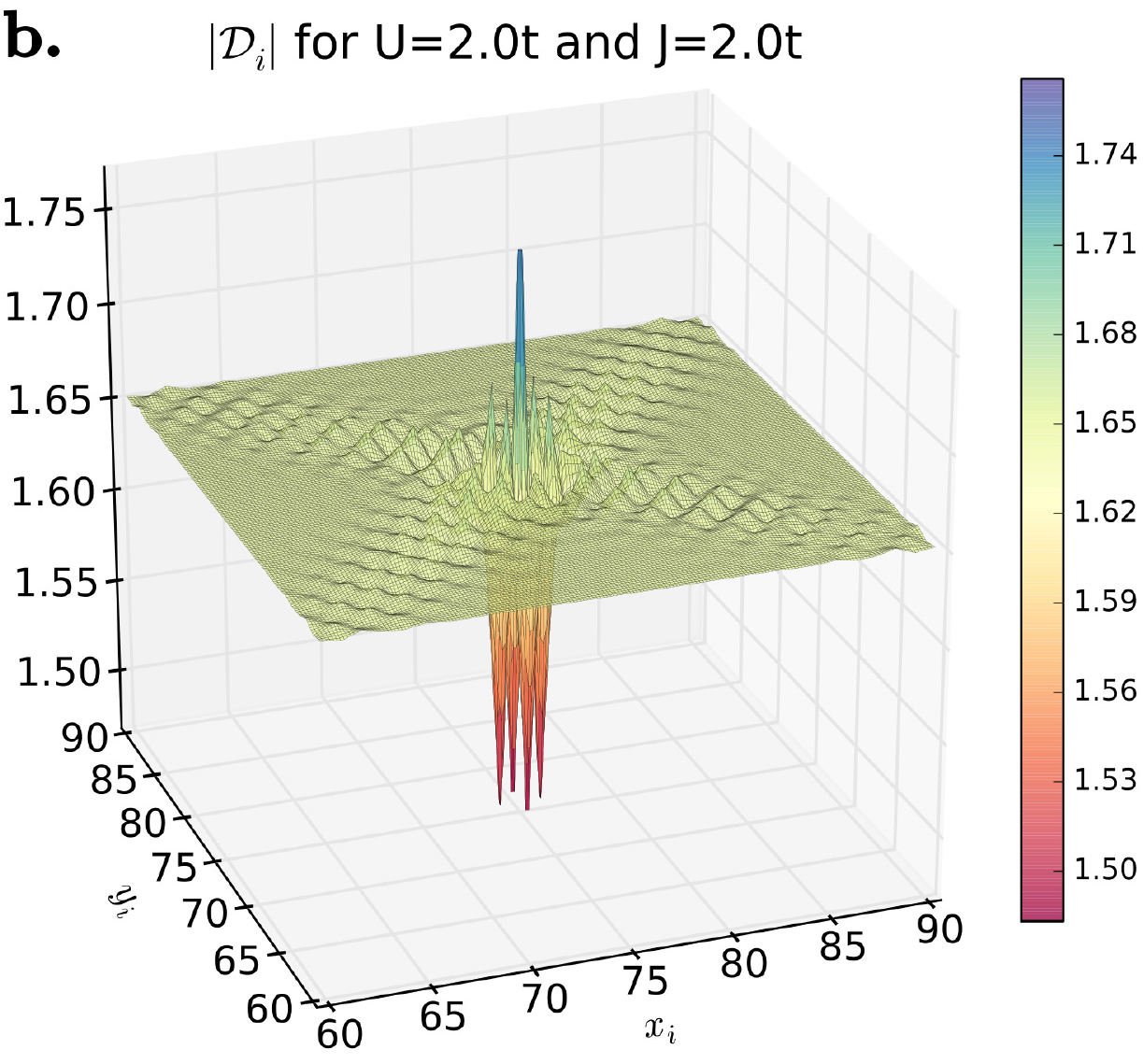}\\
\includegraphics[scale=0.35,bb=0 0 360 360]{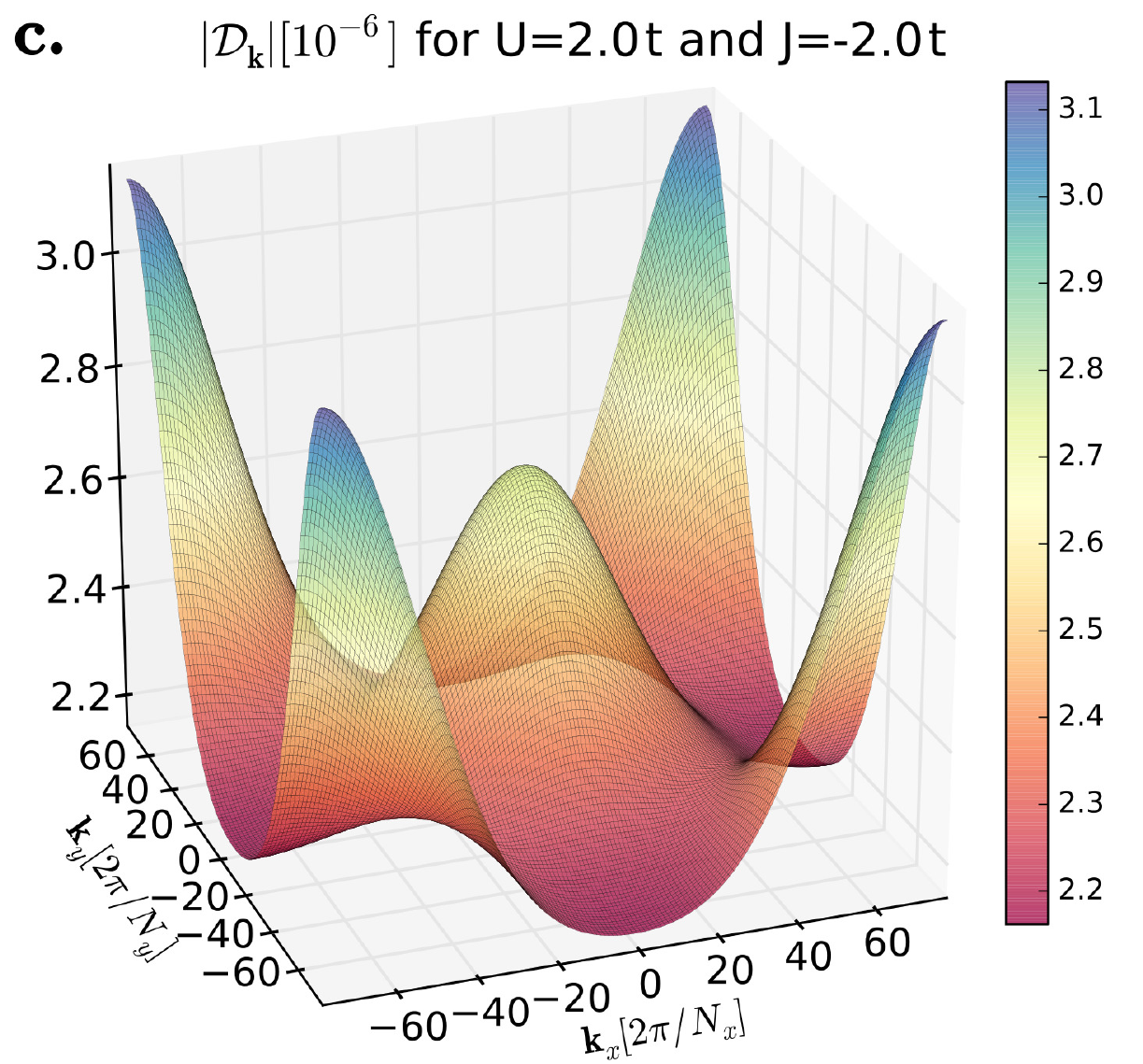}
\includegraphics[scale=0.35,bb=0 0 360 360]{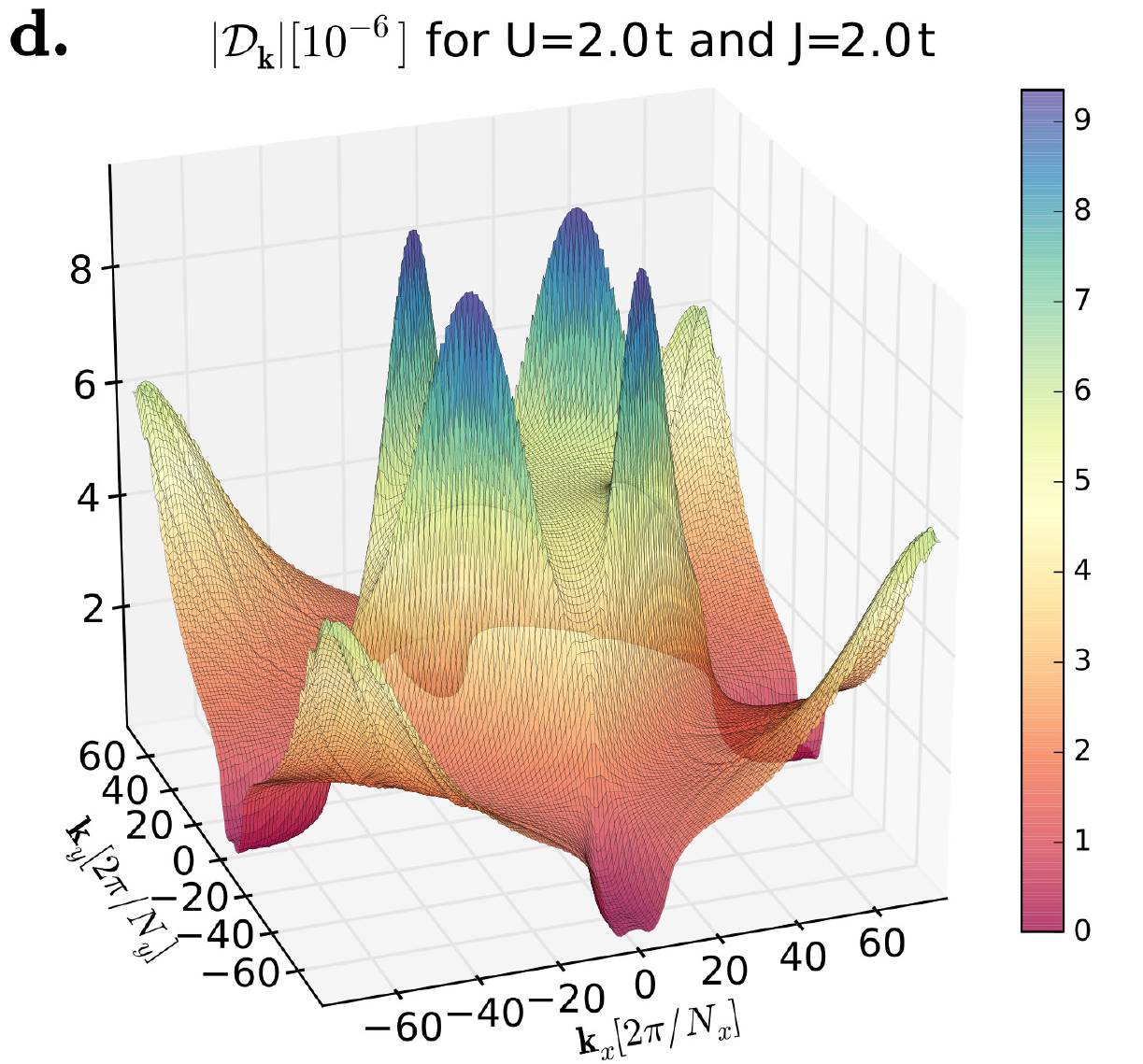}\\
\includegraphics[scale=0.35,bb=0 0 360 360]{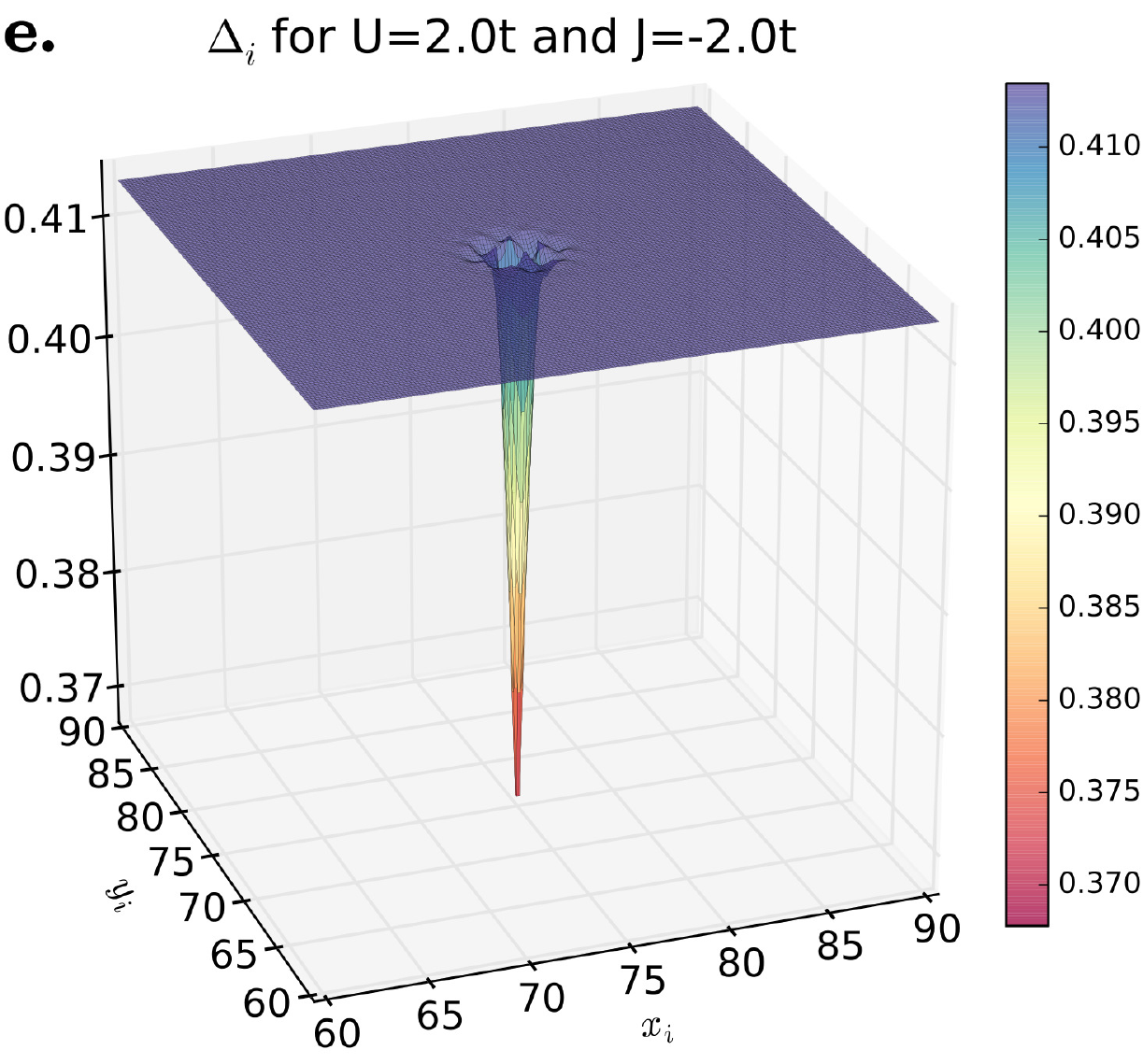}
\includegraphics[scale=0.35,bb=0 0 360 360]{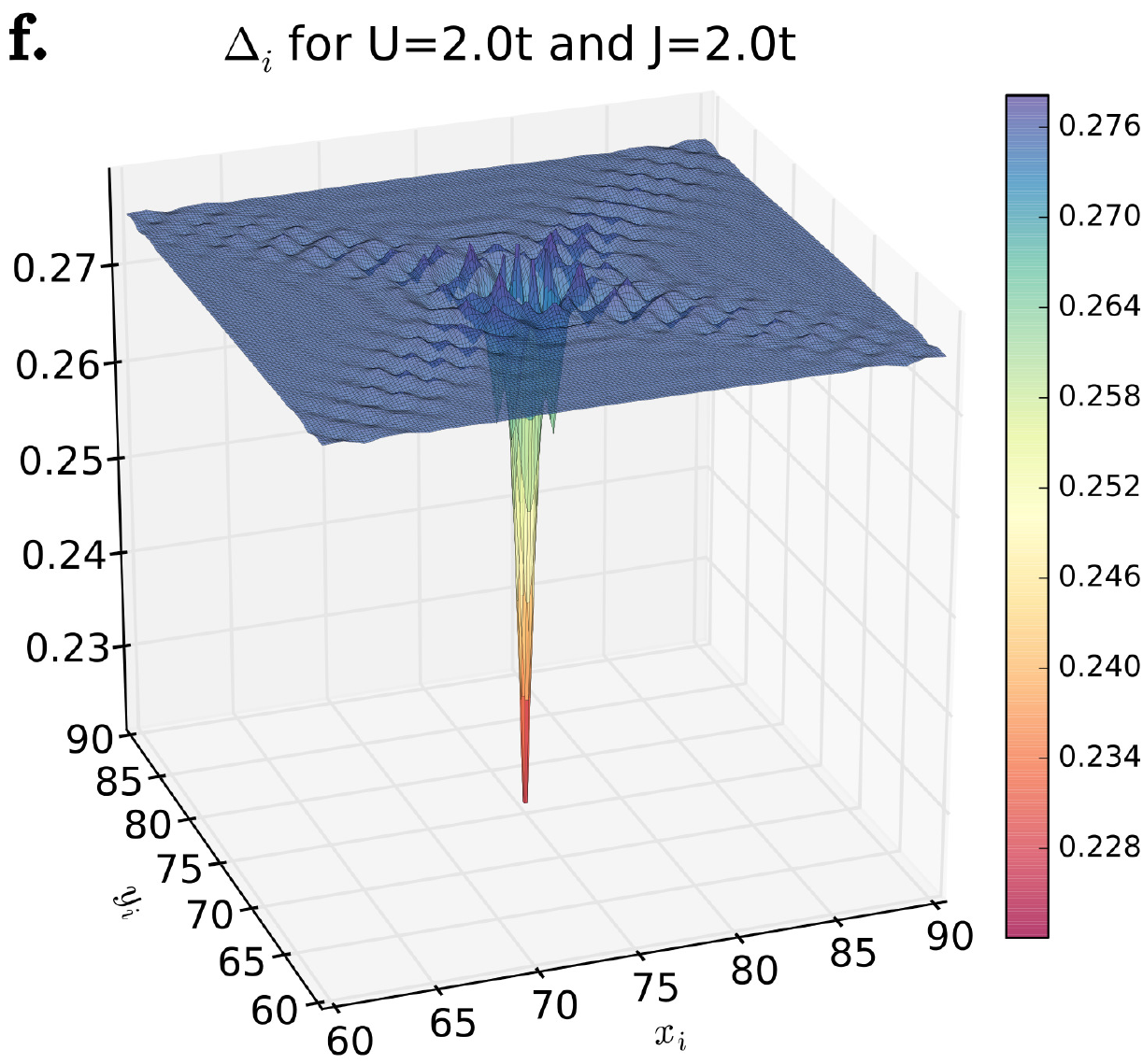}\\
\includegraphics[scale=0.35,bb=0 0 360 360]{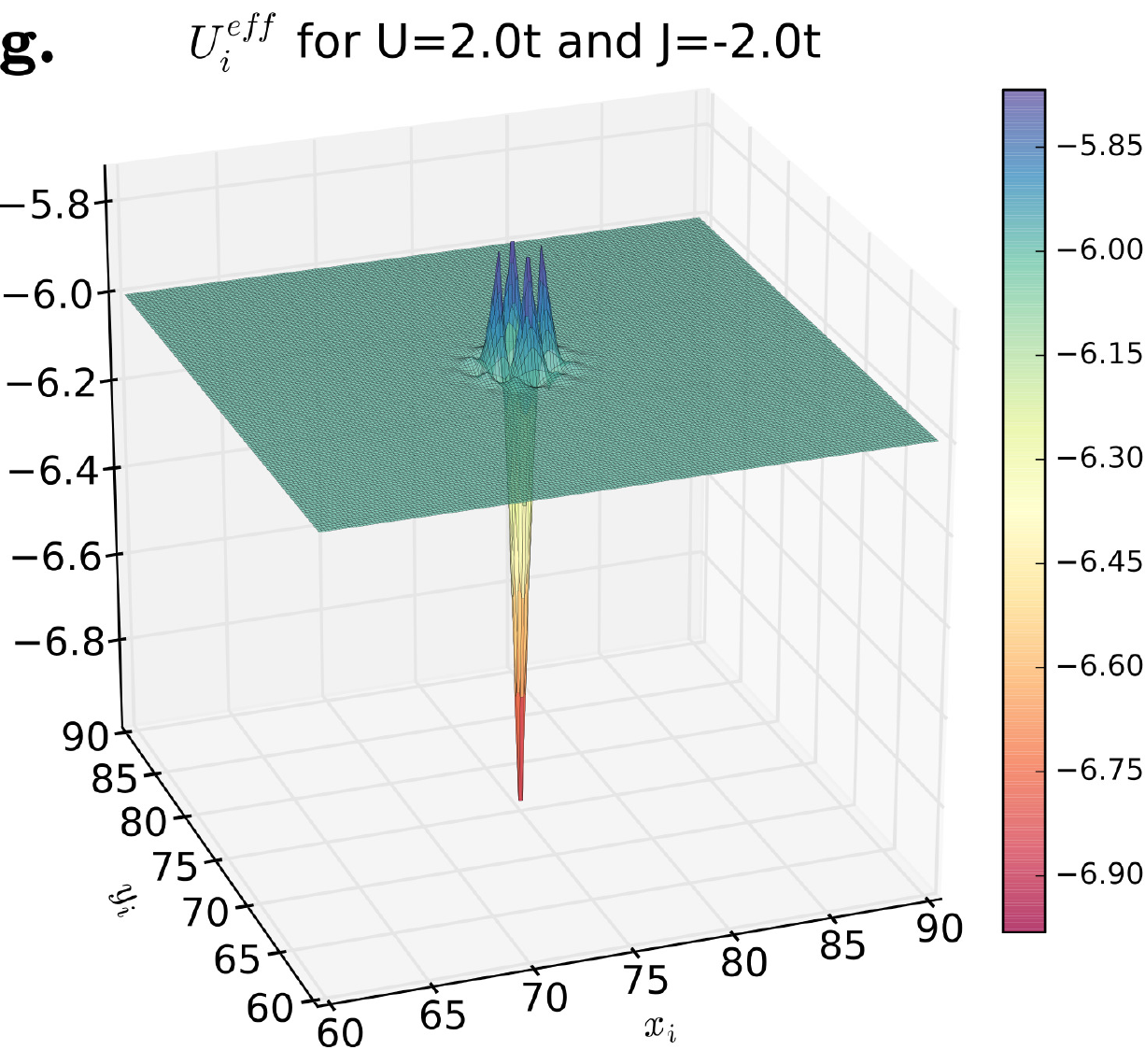}
\includegraphics[scale=0.35,bb=0 0 360 360]{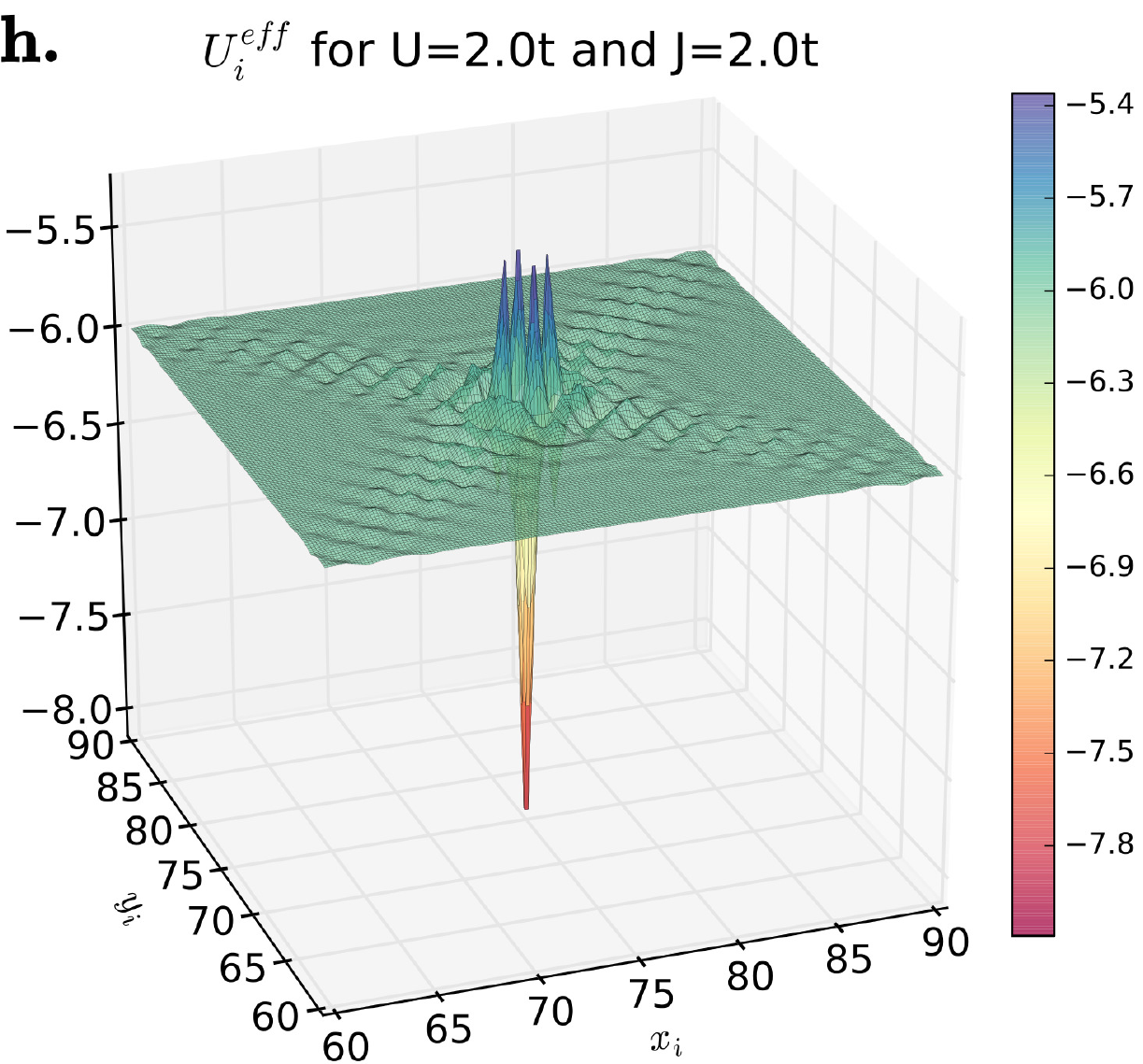}
\caption{
Plots of energy gap $|\mathcal{D}_{i}|$ (a and b), its Fourier tranform $|\mathcal{D}_{\bm k}|$ (c and d), superconducting order parameter $\Delta_{i}$ (e and f), and effective pairing interaction $U_{i}^{eff}$ (g and h) for {\it s-wave} ($U=2.0t$ and $J=-2.0t$) and $\eta${\it-wave} phase ($U=2.0t$ and $J=2.0t$) -- left and right panels, respectively. Results for the system of a size $150 \times 150$.
}\label{fig.bigg}
\end{center}
\end{figure}

From Eq.~(\ref{eq.fft}) one can obtain the distribution of the effective energy gap in the momentum space. In the homogeneous system with the SOP of {\it s} or $\eta$ type all Cooper pairs have the same momentum ${\bm Q } = {\bm 0}$ or ${\bm Q} = {\bm \Pi}$, respectively, and only $\mathcal{D}_{\bm k}$ with ${\bm k} = {\bm Q}$ are nonzero. In the presence of the impurity $V_{imp}$, the Cooper pair are scattered at the impurity site and thus in the system there are also pairs with ${\bm Q} \neq {\bm 0}, {\bm \Pi}$ (cf. Fig.~\ref{fig.bigg} panels c and d).
In particular, for the {$\eta$}{\it-wave} case, there are pairs with momentum near ${{\bm Q} = {\bm 0}}$ without distinguished direction (central peak at Fig.~\ref{fig.bigg}.d). Moreover, one can also separate pairs with momentum in  $[1,\pm1]$-direction (four peaks) located near  the central peak. These peaks are associated with long-range oscillations of gap in real space $[1,\pm1]$-direction (Fig.~\ref{fig.bigg}.b). The peaks of $\mathcal{D}_{{\bm k}}$ at $[\pm \pi, \pm \pi]$ still exist, but they are not shown in Fig.~\ref{fig.bigg}.d explicitly, because their magnitudes are much bigger.
In the {\it s-wave} phase we observe the scattering at the impurity site of the Cooper pairs with ${{\bm Q} = {\bm 0}}$ into states  with momentum near ${{\bm Q} \neq {\bm 0}}$ without distinguished direction (Fig.~\ref{fig.bigg}.c).
Moreover, in the presence of the impurity in the system, $\mathcal{D}_{{\bm k}}$ has also four peaks at  $[{\pm \pi}, {\pm \pi}]$, what means that the electrons associated with $\eta${\it-wave} superconductivity become an important component of pair states.
Notice, the peak of $\mathcal{D}_{{\bm k}}$ at $[0, 0]$ still exists, but it is not shown in Fig.~\ref{fig.bigg}.c explicitly, because its magnitude is much bigger.

It is also interesting to discuss the spatial distribution of the SOP $\Delta_i$ and effective pairing potential $U_{eff}$ in both cases, although they can not be experimentally measured in contrary to the local effective gap.
The SOP is reduced at the impurity site in the both cases (Figs.~\ref{fig.bigg}.e-f).
The behavior of $U_{eff}$ is rather unusual, because its value is enhanced at the impurity site (Figs.~\ref{fig.bigg}.g-h).
In is also worth nothing that similar long-range oscillations of these both quantities (i.e. $\Delta_i$ and $U_{eff}$ in the $[1,\pm1]$-direction in the real space (similarly as these for $\mathcal{D}_i$) are far more visible in the  $\eta$-{\it phase} than in the {\it s-wave} phase. 

The different values of the on-site impurity potential  $V_{imp}$ has been also investigated. However, the value of $V_{imp}$ has no influence on the main qualitative results of the paper.  The long-range oscillations of the energy gap (an other quantities analyzed) in real space (in the $[1,\pm1]$-direction) occurs in the {\it $\eta$-wave} phase, whereas in the {\it s-wave} phase they are absent. Only the quantitative changes are introduced by varying the value of $V_{imp}$ (the changes of magnitudes of the oscillations).

Contrasting our results with the effects of impurities on  superconducting the FFLO phase has interesting results. In this phase Cooper pairs have non-zero total momentum.
It is generally believed that that phase is very sensitive to inhomogeneities~\cite{aslamazov.68} because of the scattering of Cooper pairs on impurities.
However, some kind of disorder in the system can lead to stabilization of FFLO phase~\cite{ptok.10}.
The off-diagonal disorder can also lead to local increase of SOP at impurity.

\section{Summary and final remarks}
In this report we have analyze the influence of the nonmagnetic impurity on superconducting properties (both {\it s}- and $\eta${\it-wave}) in the materials with local electron pairing. We have shown that the scattering of Cooper pairs is into to the states which are from the neighborhoods of the states corresponding to the orderings commensurate with the crystal lattice.
Additionally, in the $\eta$-phase there are peaks in the (Fourier transform of) local superconducting gap, which are connected with long-range oscillations of the local energy gap, superconducting order parameter as well as effective pairing potential in the real-space distribution of these quantities (cf. also with Ref.~\cite{tanaka.marsiggli.2000}).
It is also interesting that the energy gap is enhanced at the impurity site for sufficiently large $J$, even if the on-site interaction $U$ is repulsive.

Notice that we do not consider the charge and magnetic orderings. They can occur for $U>0$ \cite{robaszkiewicz.bulka.99}. The more realistic analyses of the impurity impact on these states will be a topic of further works. 

The results presented in this paper can be verified, for example, by scanning tunneling microscopy (STM) technique. This technique can be used  to study impurity states in superconductors. As a first test of theories, this allows a direct comparison of local electronic features in tunneling characteristics with the theoretical predictions for the density of states and superconducting local gap (for review see e.g. Ref. \cite{balatsky.vekhter.06,krzyszczak.domanski.10} and references therein).
Thus, in correspondence with the our theoretical results for the local effective gap,  the STM spectroscopy can be useful to distinguish {\it s-wave} and $\eta${\it-wave} superconductivity in real materials. The results of this paper can give a proposal how to differentiate these two phases experimentally.

\ack
We would like to thank Yuki Nagai for suggestions about practical realization of calculation presented in Ref.~\cite{nagai.ota.12}. 
The authors thank Stanis\l{}aw Robaszkiewicz for very fruitful discussions and comments.
K.J.K. is supported by National Science Centre (Poland) -- the doctoral scholarship No.  DEC-2013/08/T/ST3/00012 and by ESF -- OP ''Human Capital'' -- POKL.04.01.01-00-133/09-00.


\section*{Bibliography}



\end{document}